\documentclass[oribibl]{llncs} 

\usepackage{float}
\usepackage{amsmath}
\usepackage{amssymb}
\usepackage{enumitem}
\usepackage{tikz-qtree}
\usepackage{url}
\usepackage{xspace}
\usepackage{subfig}

\newcommand{\vs}{\vspace{0.1cm}}

\newcommand{\cvsids}{cVSIDS\xspace}
\newcommand{\zvsids}{cVSIDS\xspace}
\newcommand{\mvsids}{mVSIDS\xspace}
\newcommand{\dvsids}{adaptVSIDS\xspace}

\newcommand{\bridgeexp}{\textit{Bridge-Experiment}\xspace}
\newcommand{\spatialexp}{\textit{Spatial-Experiment}\xspace}
\newcommand{\tempexp}{\textit{Temporal-Experiment}\xspace}

\let\doendproof\endproof
\renewcommand\endproof{~\hfill\qed\doendproof}

\title{Understanding VSIDS Branching Heuristics\\ in Conflict-Driven Clause-Learning SAT Solvers}

\author{
Jia Hui Liang \and 
Vijay Ganesh \and
Ed Zulkoski \and \\
Atulan Zaman \and
Krzysztof Czarnecki
}

\institute{University of Waterloo, Waterloo, Canada}

\begin{document}
\maketitle

\begin{abstract}

Conflict-Driven Clause-Learning (CDCL) SAT solvers crucially depend on
the Variable State Independent Decaying Sum (VSIDS) branching
heuristic for their performance.  Although VSIDS was proposed nearly
fifteen years ago, and many other branching heuristics for SAT solving
have since been proposed, VSIDS remains one of the most effective
branching heuristics. Despite its widespread use and repeated attempts
to understand it, this \emph{additive bumping} and
\emph{multiplicative decay} branching heuristic has remained an enigma.

In this paper, we advance our understanding of VSIDS by answering the
following key questions. The first question we pose is ``what is
special about the class of variables that VSIDS chooses to additively
bump?'' In answering this question we showed that VSIDS overwhelmingly
picks, bumps, and learns bridge variables, defined as the variables
that connect distinct communities in the community structure of
SAT instances. This is surprising since VSIDS was invented more than a
decade before the link between community structure and SAT solver
performance was discovered. Additionally, we show that VSIDS viewed as
a ranking function correlates strongly with temporal graph centrality
measures. Putting these two findings together, we conclude that VSIDS
picks high-centrality bridge variables. The second
question we pose is ``what role does multiplicative decay play in
making VSIDS so effective?'' We show that the multiplicative decay
behaves like an exponential moving average (EMA) that favors variables
that persistently occur in conflicts (the signal) over variables that
occur intermittently (the noise). The third question we pose is
``whether VSIDS is temporally and spatially focused.'' We show that
VSIDS disproportionately picks variables from a few communities
unlike, say, the random branching heuristic. We put these findings
together to invent a new adaptive VSIDS branching heuristic that
solves more instances than one of the best-known VSIDS variants over
the SAT Competition 2013 benchmarks.
\end{abstract}

\section{Introduction}

The Boolean satisfiability (SAT) problem~\cite{npcook} is the
quintessential NP-complete
problem, a class of decision
problems conjectured to be computationally hard. Yet, impressively,
modern sequential Conflict-Driven Clause-Learning SAT
solvers~\cite{minisat,lingeling,zchaff,grasp,glucose} are able to
solve large instances obtained from real-world
applications~\cite{empircalsat, satcomp}.  Although hundreds of
techniques and heuristics have been proposed over the last five
decades to solve the Boolean SAT problem~\cite{satconf,satcomp},
modern SAT solvers rely crucially only on a handful of them.  Of
these, the two most important are Conflict-Driven Clause-Learning with
backjumping (CDCL)~\cite{grasp} and Variable State Independent
Decaying Sum (VSIDS) branching heuristic~\cite{chaff}.  Many
systematic experiments have been performed to ascertain the veracity
of this observation~\cite{empircalsat}.  Additionally, not only is
VSIDS one of the most effective branching heuristics, but many other
well-known high-performing branching heuristics are simply variants of
VSIDS. Researchers have proposed some theoretical explanations for the
impact of clause-learning on the performance of the modern SAT
solvers: clause-learning allows SAT solvers to polynomially simulate
general resolution propositional proof
system~\cite{knot,atserias,beame}.  However, our understanding of the
role played by VSIDS heuristic has previously been limited. The motivation
for the research presented in this paper is to achieve a better
scientific understanding of VSIDS. We focus on two
well-known variations of VSIDS, namely \cvsids and \mvsids, described
in Section~\ref{sec:background}.


\vs
\noindent{\bf Our Scientific Findings and Contributions.} 
In this paper we ask the following questions regarding the behavior of
VSIDS.\footnote{All code and experimental data sets are available from
our website: \url{https://github.com/JLiangWaterloo/vsids}.} First,
what is special about the class of variables that VSIDS chooses to
additively bump? (Answered by Contributions I and III.) Second, what
role does multiplicative decay play in making VSIDS so effective?
(Answered by Contribution IV.) Third, is VSIDS temporally and
spatially focused? (Answered by Contribution II.)

\vs
\noindent{\bf Contribution I: Bridge Variables and VSIDS.} 
Community structure is a property exhibited in many real-world graphs,
particularly in social networks, where the graph can be partitioned
into groups of vertices, called communities, such that each group is
densely connected within itself but sparsely connected with other
groups. Recent research has shown that the community
structure quality of the SAT input correlates with faster solving
time~\cite{satcommunitypaper}.  We show that bridge variables
connecting distinct communities in the community structure of a SAT
instance~\cite{community} are high priority targets for both the
branching heuristic and clause-learning, which suggests one possible
explanation for this correlation.

\vs
\noindent{\bf Contribution II: Community-focused Search and VSIDS.}
We define two terms, \textit{spatial focus} and \textit{temporal
  focus}, to describe how a branching heuristic focuses on certain
regions of the search space during solving, with respect to the
underlying community structure. We
refer to this form of locality as focused search, to distinguish it
from local search performed by stochastic local search
solvers~\cite{sls}. We show that \mvsids is more focused
than \cvsids and random branching according to these metrics.

\vs
\noindent{\bf Contribution III: Exponentially-smoothed Temporal Graph
  Centrality and VSIDS correlate strongly.} Third, we show that VSIDS
rankings correlate strongly with the variable rankings induced by
\emph{exponentially smoothed temporal graph centrality} (TGC) measures
over the \emph{temporal variable incidence graphs} (TVIG) of the
original and learnt clauses of an input SAT instance. This correlation
remains strong throughout the run of the solver. The TVIG extends the
well-known \emph{variable incidence graph} over Boolean formulas by
incorporating the dynamically evolving aspect of the learnt clause
database inside a SAT solver and uses exponential smoothing to focus
on recently learnt clauses.  TGC is the temporal version of the
widely-used graph centrality measures, such as degree and eigenvector
centrality, which are used to identify important vertices in a graph.
The definitions are inspired by recent research on temporal aspects of
social networks~\cite{temporalsocialnetwork1,timedpagerank}.  For
example, the timed PageRank algorithm~\cite{timedpagerank} is used to
discover important publications that are likely to be referenced in
the future.  We show that VSIDS typically selects variables with high
\emph{temporal degree centrality} and \emph{temporal eigenvector
  centrality}. The above-mentioned findings essentially tell us that
we have a single family of mathematically-precise graph-theoretic
measures, namely TGC, that succinctly characterizes both the additive
bump and multiplicative decay components of VSIDS family of
heuristics. Variables that have high centrality correspond to
variables in ``recent'' learnt clauses that are ``highly-constrained''
and get additively bumped.  Variables that are not ``persistently''
highly-constrained, i.e., do not occur frequently in recent learnt
clauses get decayed away quickly. Putting together Contributions I and
III, we conclude that {\it VSIDS picks
  high-centrality bridge variables}.

\vs
\noindent{\bf Contribution IV: Exponential Moving Average and
  Multiplicative Decay in VSIDS.}  Fourth, we show that the
multiplicative decay in VSIDS is a form of exponential moving average,
and provide a plausible explanation as to why this is crucial to the
effectiveness of VSIDS.

\vs
\noindent{\bf Contribution V: A Novel Adaptive Branching Heuristic.}
Our findings led to a new VSIDS called \dvsids that adapatively
adjusts the exponential moving average (a form of adaptive
moving average) depending on the quality of the learnt clauses. We
show that \dvsids outperforms \mvsids, by
solving 2.4\% more instances over the SAT Competition
2013 benchmarks.

\section{Background}
\label{sec:background}
Here we describe VSIDS and the variable incidence graph of a CNF
formula.

\vs
\noindent{\bf The VSIDS Branching Heuristic and Variants.}
The term VSIDS refers to a family of branching heuristics widely used
in modern CDCL SAT solvers that \emph{rank} all variables of a Boolean
formula during the run of a solver. 
As things
stand today, VSIDS is significantly more effective than other
well-known heuristics such as DLIS~\cite{dlis}, MOM~\cite{mom},
Jeroslow-Wang~\cite{jeroslowwang}, and BOHM~\cite{bohm}. VSIDS was a
major breakthrough when first introduced as part of the Chaff
solver~\cite{chaff}. The key idea is to collect statistics over learnt
clauses to guide the direction of the search, where recent learnt
clauses are favored.
The key characteristics of VSIDS
is the additive bumping and multiplicative decay behavior, described
in more details below. Another positive characteristic of VSIDS is its low computational
overhead. We focus on two of the more well-known variants
of VSIDS, namely, the variant from Chaff~\cite{chaff} and the variant
from MiniSAT version 2.2.0~\cite{minisat}.  We refer to these variants
as \cvsids and \mvsids respectively.  Both variants have the common
characteristics listed below.

\vs
\noindent{\bf Activity Score, Initialization and VSIDS Ranking.}
VSIDS assigns a floating point number, called \emph{activity}, to each
variable in the Boolean formula. At the begining of a run of a solver,
the activity scores of all variables are typically initialized to
0. We refer to the ranking of variables according to their activity
scores in the decreasing order as the VSIDS ranking. VSIDS picks the
variable with the highest activity to branch on.
 
\noindent{\bf Additive Bump and Multiplicative Decay.}
When the solver learns a clause, a set of variables is chosen, and
their activities are additively increased, typically by $1$.  The
quantum of this increase is called the (additive) \emph{bump}. At
regular intervals during the run of the solver, the activities of all
variables are multiplied by a constant $0 < \alpha < 1$ called the
(multiplicative) \emph{decay factor}.

\vs
\noindent{\bf \zvsids.} The activities of variables occurring in the
newest learnt clause are bumped up by $1$, immediately after the
clause is learnt.  The activities of all variables are multiplied by a
constant $0 < \alpha < 1$.  The decay occurs after every $i$
conflicts.  We follow the policy used in recent solvers like
MiniSAT and use $i=1$.

\noindent{\bf \mvsids.} The activities of all variables resolved
during conflict analysis that lead to the learnt clause (including the
variables in the learnt clause) are bumped up by $1$.  The activities
of all variables are decayed as in \cvsids~\footnote{MiniSAT's actual
implementation is slightly different, but has the same effect. Rather
than decaying the activities of every variable, it increases the bump
quantum of all future conflicts instead~\cite{armin2008}.}.

\vs
\noindent{\bf Variable Incidence Graph (VIG).}
The VIG of a CNF formula $F$ is defined as follows: vertices of the
graph are the variables in the formula.  For every clause $c \in F$ we
have an edge between each pair of variables in $c$.  In other words,
each clause corresponds to a clique between its variables.  The weight
of an edge is $\frac{1}{|c|-1}$ where $|c|$ is the length of the
clause.  VIG does not distinguish between positive and negative
occurrences of variables.  We combine all edges between each pair of
vertices into one weighted edge by summing the weights. More precisely,
the VIG of a CNF formula $F$ is a weighted graph defined as follows:
set of vertices $V = Var$, set of edges $E =\{xy \mid x,y\in
c\in F\}$, and the weight function $w(xy) = \sum_{x,y\in c
  \in F} \frac{1}{|c|-1}$.

\section{Contribution I and II: Community-focused Search, Bridge Variables, and VSIDS}
\label{sec:focus}

In this section, we describe the experimental setup, methodology, and
results to show the connection between VSIDS and community structure.

\vs
\noindent{\bf The Hypotheses.} Here we state the three hypotheses that
we tested in this section: {\bf 1) Bridge Experiment:} VSIDS
disproportionately picks, bumps, and learns the bridge variables, {\bf
  2) Spatial Focus Experiment:} VSIDS disproportionately picks from a
smaller number of communities rather than a large fraction of the
communities of a SAT instance, and {\bf 3) Temporal Focus Experiment:}
VSIDS typically picks from recently-seen communities.

\vs
\noindent{\bf Community Structure of the Graph of SAT Instances, and
  Bridge Variables.} The concept of decomposing graphs into {\it
  natural communities}~\cite{clauset2004finding,zhang2013online} arose
in the study of complex networks such as the {\it graph of biological
  systems}. Informally, a network or graph is said to have community
structure if the graph can be decomposed into sub-graphs where the
sub-graphs have more internal edges than outgoing
edges~\cite{satcommunitypaper}. We say that a graph has a ``good''
community structure if the percentage of intra-community edges is
significantly higher than inter-community edges. We refer to these
inter-community edges as bridges, and the vertices connected
by such edges as {\it bridge vertices}. In the context of the
community structure of the VIG of a Boolean formula, bridge vertices
are called {\it bridge variables}. We refer the reader to these
papers~\cite{clauset2004finding,zhang2013online} for a more formal
introduction to community structure of graphs.

Recently there has been some interesting discoveries regarding the
impact of community on CDCL SAT solver
performance~\cite{satcommunitypaper}. Specifically, the authors of the
paper~\cite{satcommunitypaper} showed that the running time of CDCL
solvers is strongly correlated with community structures of SAT
instances. In light of these discoveries, it was but natural for us
ask the question whether VSIDS somehow exploits the community
structure of SAT instances. What we discovered and explain below is
that VSIDS disproportionately picks, bumps, and learns the bridge
variables in the community structure of SAT instances.

\vs
\noindent{\bf Temporal and Spatial Focused Search.} We further define
two terms, \textit{spatial focus} and \textit{temporal focus}, to
describe how a branching heuristic gravitates towards certain regions
of the search space during solving, with respect to the underlying
community structure.  We say a branching heuristic is spatially
focused if it disproportionately picks variables from a small set of
communities, when normalized for size, throughout the entire
run of the solver.  A branching heuristic exhibits temporal focus if
it typically picks a new decision variable from a small fixed-size
\textit{window} of recently-seen communities.

\vs
\noindent{\bf Experimental Setup and Methodology.} Experiments were
performed over the 1030 instances from SAT Competition
2013~\cite{satcomp}, after simplification using MiniSAT
simplifying-solver. We use the Louvain method~\cite{louvain} to
compute the communities of the VIG of the input SAT formulas. There
are many community-detecting algorithms to choose from and we picked
Louvain because it scales well with the size of input graphs. For each
instance, the Louvain method is given an hour to compute and save the
communities it finds. The community information is then given to a
modified MiniSAT 2.2.0 so it can track the bridge variables. Due to the high cost,
we only compute the communities once at the start.

For the \bridgeexp, we ran the instances using a modified MiniSAT with
a timeout of 5000 seconds, as per the SAT Competition 2013 rules.
Before MiniSAT begins its CDCL loop, it reads in the community
information stored by the Louvain method. The solver then scans
through its the initial input clauses and checks which variables share
at least one clause with another variable residing in a different
community and marks them as bridge variables. Whenever our modified
version of MiniSAT 1) picks a decision variable, 2) bumps a variable,
and 3) learns a clause over a variable during the search, it checks
whether the variable is a bridge variable. If so, the solver updates
its internal counters to keep track of the number of bridge variables
in the each of the 3 scenarios. At the end of the run, the solver
outputs the percentage of variables that are bridge in each of these
scenarios. This additional code adds little overhead and does not
change the behavior of MiniSAT. We are simply instrumenting the solver
to collect statistics of interest.  For the \tempexp and \spatialexp,
we additionally modified MiniSAT to record all decision variables to a
file, in order to post-process the data. We allowed a 10000 second
timeout for these experiments due to this additional overhead.

\vs
\noindent{\bf The Reporting of Results.} In the \bridgeexp, for each
instance, we compute the percentage of decision variables, bumped
variables, learnt clause variables, and number of variables that are
also bridges. Then we averaged these percentages over the three SAT
2013 Competition benchmark categories (application,
combinatorial, and random) and
reported these numbers.

For the \spatialexp, for every community $i$, we compute a community
score $cs_i=picks\_from(i)/order(i)$, where $picks\_from(i)$ is the number of times the solver branched on a variable from community $i$ and $order(i)$ is the size of community $i$ in terms of variables. We then use the
\textit{Gini coefficient}~\cite{gini}, a statistical measure of
inequality, to compute our spatial score $ss = gini(cs_i \text{ for i
}\in communities)$. A score of 1 indicates total disparity (e.g. all
picks are from one community), whereas zero indicates total
equality. Higher scores therefore favor our hypothesis.  We report the
average $ss$ value for each benchmark category. The intuition behind this experiment and
the use of the Gini coefficient here (used in measuring the inequality
of wealth distribution in countries) is that it is an effective method
for computing how unequally a branching heuristic favors some
communities over others. Using this metric we show for example that
VSIDS disproportionately favors a small set of communities (highly
unequal distribution of picks) versus random branching heuristic
(largely equal distribution of picks).

For the \tempexp, we define our window size $ws$ to be 10\% of the
total number of communities, rounded up to the nearest integer. For
all instances, our window contains the set of communities
from the $ws$ most recent decisions (note that the set may have less
than $ws$ elements). At every decision, we increment a counter
$window\_hits$ if the current variable is from a community in the
window. We assign a temporal score $ts = window\_hits / decisions$ for
each instance.  We report the average $ts$ value for each benchmark
category. The key idea behind this experiment is to test the
hypothesis that VSIDS branching favors picking from recently picked-from communities, versus random which does not display such temporal locality.

\vs
\noindent{\bf Results and Interpretations of Bridge Variable
  Experiment.}  Table~\ref{table:bridge} shows that bridge variables
are highly favored in MiniSAT by its branching heuristic, conflict
analysis, and clause-learning. It is a surprising result that bridge
variables are favored even though the heuristics and techniques in
MiniSAT have no notion of communities.  While bridge variables
certainly make up a large percent of variables, the percent of picked
bridge variables is even higher.
Table \ref{table:bridge} includes only the instances where the Louvain
implementation completed before timing out. In total, 229/300
instances in the application category and 238/300 instances in the
hard combinatorial category are included in the
Table~\ref{table:bridge}. In the random category, every variable is a
bridge, hence the results are omitted. This is expected because it is
highly improbable to generate random instances where a variable is not
neighboring another variable outside its community.

\begin{table}[t]
	\centering
	\scalebox{0.8}{
		\begin{tabular}{ | l c c c c | }
			\hline
			Category
			& \quad\begin{tabular}[x]{@{}c@{}}\% of\\variables\\that are bridge\end{tabular}\quad
			& \quad\begin{tabular}[x]{@{}c@{}}\% of\\picked variables\\that are bridge\end{tabular}\quad
			& \quad\begin{tabular}[x]{@{}c@{}}\% of\\bumped variables\\that are bridge\end{tabular}\quad
			& \quad\begin{tabular}[x]{@{}c@{}}\% of\\learnt clause variables\\that are bridge\end{tabular}\quad \\
			\hline
			Application & 61.0 & 79.9 & 71.6 & 78.4 \\
			Combinatorial & 78.2 & 87.6 & 84.3 & 88.2 \\
			\hline
		\end{tabular}
	}
	\caption{MiniSAT's CDCL and \mvsids techniques prefers to pick, bump, and learn
		over bridge variables.}
	\label{table:bridge}
	\vspace{-0.25in}
\end{table}
\begin{table}[t]
        \centering
        \subfloat[\spatialexp average $ss$ score.]{
                \begin{tabular}{ | l | c c c | }
                        \hline
                        Category & \mvsids & \cvsids & random \\
                        \hline
                        Application & 0.592 & 0.560 & 0.216 \\
                        Combinatorial & 0.275 & 0.261 & 0.099 \\
                        Random  & 0.029 & 0.023 & 0.006 \\
                        \hline
                \end{tabular}
        }
        \subfloat[\tempexp average $ts$ score.]{
                \begin{tabular}{ | l | c c c | }
                        \hline
                        Category & \mvsids & \cvsids & random \\
                        \hline
                        Application & 0.580 & 0.551 & 0.268 \\
                        Combinatorial & 0.505 & 0.473 & 0.265 \\
                        Random  & 0.269 & 0.268 & 0.219 \\
                        \hline
                \end{tabular}
                
        }
        \caption{(a) VSIDS heuristics are more spatially focused than random branching. (b) VSIDS heuristics tend to pick from recently-picked communities.}
        \label{table:focus}
        \vspace{-0.25in}
\end{table}

Recent research suggests that CDCL solvers take advantage of good
community structure in SAT instances~\cite{satcommunitypaper} leading
to faster solving time. The reason for this phenomenon is not fully
understood. One possibility is that good community structure lends
itself to divide-and-conquer because the bridges are easier to cut
(i.e., satisfy). More precisely, the
solver can focus its attention on the bridges by picking the bridge
variables and assigning them appropriate values.  When it eventually
assigns the correct values to enough bridges, the VIG is divided into
multiple components, and each component can be solved with no
interference from each other. Even if the VIG cannot be completely
separated, it may still be beneficial to the cut bridges between
communities so that these communities can be solved relatively
independently.

\vs
\noindent{\bf Results and Interpretations of Temporal and Spatial
  Focused Search Experiments.}  Table \ref{table:focus}a depicts the
average Gini coefficient for the \spatialexp. Both VSIDS techniques
exhibit much more inequality relative to random branching for the
application and combinatorial instances, indicating that VSIDS may be
attempting to \textit{hone in} on certain communities. The very low
values for random instances indicate that none of the branching
heuristics typically favor certain communities, likely due to the poor
community structures exhibited by such instances. Table
\ref{table:focus}b demonstrates that VSIDS techniques are much more
temporally focused on average than random branching. It is
commonly believed that VSIDS improves the \emph{search
  locality}~\cite{zchaff,nadel2010assignment} which in turn improves
solver performance. However, this term \emph{search locality} has
previously been not rigorously defined. We precisely defined spatial
focus and temporal focus, and show that VSIDS displays high search
locality in terms of these definitions.

\section{Contribution III: Experimental Evidence Supporting Strong Correlation Between TGC and VSIDS}
\label{sec:resultstgc}

In this section, we describe the experiments to support the hypothesis
that the VSIDS variants \cvsids and \mvsids, viewed as ranking
functions, correlate strongly with both temporal degree centrality and
temporal eigenvector centrality according to Spearman's rank
correlation coefficient and top-k measures. Combining the results of
this section with Contribution I (namely, VSIDS picks, bumps and
learns over bridge variables), we conclude that VSIDS picks
{\it high-centrality bridge variables}.

\vs
\noindent{\bf Temporal Variable Incidence Graph (TVIG).} To incorporate the temporal aspect of learnt clauses
we introduce \emph{temporal variable incidence graph} (TVIG) here,
that extends the VIG by encoding temporal information into its
structure. In the TVIG, every clause is labeled with a timestamp
denoted $t(c)$.  The $t(c)$ is equal to $0$ if $c$ is a clause from
the original input formula, otherwise
$t(c)$ is equal to the number conflicts up to the learning of $c$. We
refer to the difference between the current time $t$ and the timestamp
of a clause $t(c)$ as the age of the clause: $age(c) = t-t(c)$.  Fix
an \emph{exponential smoothing factor} $0<\alpha<1$. The TVIG is a
weighted graph constructed in the same manner as the VIG except the
weight of an edge is $\frac{\alpha^{age(e)}}{|c|-1}$. Like the VIG,
multiple edges between a pair of vertices are combined into one
weighted edge.  More precisely, the TVIG of a clause database at time
$t$ is defined in the same way as VIG except with a modified weight
function that takes the ages of clauses into account: $w(xy) =
\sum_{x,y\in c \in F} \frac{\alpha^{age(c)}}{|c|-1}$. Observe that the
TVIG evolves throughout the solving process: as new learnt clauses are
added, new edges are added to the graph, and all the ages increase.
As an edge's age increases, its weight decreases exponentially with
time assuming no new learnt clause contains its variables.
In many domains, it is often the case that
more recent data points are more useful than older data points.

\vs
\noindent{\bf (Temporal) Degree and Eigenvector Centrality.}  A graph
centrality measure is a function that assigns a real number to each
vertex in a graph.  The number associated with each vertex denotes its
relative importance in the
graph~\cite{centralityfreeman1979,centralitysocial,gouldindexjustification}.
For example, the degree centrality~\cite{centralitysocial} of a vertex in a
graph is defined as the degree of the vertex. The eigenvector
centrality of a vertex in a graph is defined as its corresponding
value in the eigenvector of the greatest eigenvalue of the graph's
adjacency matrix. We similarly define the temporal versions of degree
and eigenvector centrality. The key idea needed to define temporal
graph centrality measures is to incorporate temporal information
inside the TVIG.  The temporal degree centrality (TDC) and (resp. temporal eigenvector centrality (TEC)) of a vertex at
time $t$ is defined as the degree centrality (resp. eigenvector centrality) of the vertex in the TVIG
at time $t$.

\vs
\noindent{\bf Experimental Setup and Methodology.}  We implemented the
VSIDS variants and TGC measures in MiniSAT 2.2.0~\cite{minisat}. All
the experiments were performed using MiniSAT on all 1030 Boolean formulas obtained from all
three categories (application, combinatorial, and random) of the
SAT Competition 2013~\cite{satcomp}. Before beginning any experimentation, the instances are first
simplified using MiniSAT's inbuilt preprocessor with the default
settings. All experiments were
performed on the SHARCNET cloud~\cite{sharcnet}, where cores range in
specs between 2.2 to 2.7 GHz with 4 GB of memory, and 4 hour
timeout. We use 100 iterations of the power iteration
algorithm~\cite{poweriteration} to compute TEC, and 1 iteration for
TDC.  We use MiniSAT's default decay factor of $0.95$ for
VSIDS. We also use $0.95$ as the exponential smoothing
factor for the TVIG. We take measurements on the
current state of the solver after every 5000 iterations, where an
iteration is defined as a decision or a conflict.
Observe that we take measurements
dynamically as the solver solves an instance, and not just once at the
beginning. Such a dynamic comparison gives us a much better picture of
the correlation between two different ranking functions or measures
than a single point of comparison.

\vs
\noindent{\bf Methodology for Comparing Rankings based on Spearman's
  Rank Correlation Coefficient.}  For each set of experiments, for
each SAT instance, for every measurement made, we compute the
Spearman's rank correlation coefficient~\cite{spearman} between the
VSIDS and TGC rankings. Spearman's rank correlation coefficient is a
widely-used correlation coefficient in statistics for measuring the
degree of relationship between a pair of rankings. The strength of
Spearman's correlation is conventionally interpreted as
follows: 0.00--0.19 is very weak, 0.20--0.39 is weak, 0.40--0.59 is
moderate, 0.60--0.79 is strong, 0.80--1.00 is very strong.  We compute
the average of the Spearman's correlation over the
execution of a SAT solver on each instance.  We follow the standard
practice of applying the Fisher
transformation~\cite{fishertransformation} when aggregating the
correlations.

\vs
\noindent{\bf Methodology for Comparing Rankings based on Top-k.}
Let $v$ be the unassigned variable with the highest ranked according to some VSIDS
variant.  Let $i$ be the position of variable $v$ according to a
specific TGC ranking, excluding assigned variables. Then the
\emph{top-$k$ measure} is $1$ if $i\leq k$, otherwise $0$.
The rationale
for this metric is that SAT solvers typically only choose the
top-ranked unassigned variable, according to the VSIDS ranking, to
branch on. If
the VSIDS top-ranked unassigned variable occurs very often among the
top-k ranked variables according to TGC, then we infer that VSIDS
picks variables that are highly ranked according to TGC. In our
experiments, we used various values for $k$. Again, we compute the
average of top-k measure over the execution of a SAT solver on each
instance.

\vs
\noindent{\bf The Reporting of Results.}  For every pair of rankings,
one from the VSIDS family and the other from the TGC family, we report the
top-k measure and Spearman's rank correlation coefficient between the
pair of rankings every 5000 iterations. On termination, we compute the
average for the instance. We take all the instance averages and
average them again, and report the average of the averages.  The final
numbers are labeled as ``mean top-k" or ``mean Spearman". For example,
a mean top-10 of 0.912 is interpreted as ``for the average instance in
the experiment, 91.2\% of the measured top-ranked variables according
to VSIDS are among the 10 unassigned variables with the highest
centrality". Likewise, a high mean Spearman implies the average
instance has a strong positive correlation between VSIDS and TGC
rankings.

\vs
\noindent{\bf Results and Interpretations.}  In Table~\ref{table:tdc}
(resp. Table~\ref{table:tec}), we compare VSIDS and TDC (resp. TEC)
rankings. The data shows a strong correlation between VSIDS and TDC,
in particular, the 0.818 mean Spearman
between \cvsids and TDC is high. The metrics are lower with TEC, but
the correlation remains strong. 
\mvsids has a better mean Spearman with TEC
than TDC in the application category. 
We have also conducted this experiment with non-temporal
degree/eigenvector centrality and the resulting
mean Spearman and mean top-k are significantly
lower than their temporal counterparts.

\begin{table}[t]
\centering
\scalebox{0.8}{
\begin{tabular}{ | l | c c c | c c c | }
\hline
& \multicolumn{3}{|c|}{\cvsids vs TDC} & \multicolumn{3}{|c|}{\mvsids vs TDC} \\
& Application & \quad Combinatorial \quad & Random & Application & \quad Combinatorial \quad & Random \\ 
\hline
Mean Spearman & 0.818 & 0.946 & 0.988 & 0.629 & 0.791 & 0.864 \\
Mean Top-1    & 0.884 & 0.865 & 0.949 & 0.427 & 0.391 & 0.469 \\
Mean Top-10   & 0.912 & 0.898 & 0.981 & 0.705 & 0.735 & 0.867 \\
\hline
\end{tabular}
}
\caption{Results of comparing VSIDS and TDC.}
\label{table:tdc}
\vspace{-0.25in}
\end{table}

\begin{table}[t]
\centering
\scalebox{0.8}{
\begin{tabular}{ | l | c c c | c c c | }
\hline
& \multicolumn{3}{|c|}{\cvsids vs TEC} & \multicolumn{3}{|c|}{\mvsids vs TEC} \\
& Application & \quad Combinatorial \quad & Random & Application & \quad Combinatorial \quad & Random \\ 
\hline
Mean Spearman & 0.790 & 0.926 & 0.987 & 0.675 & 0.764 & 0.863 \\
Mean Top-1    & 0.470 & 0.526 & 0.794 & 0.293 & 0.304 & 0.418 \\
Mean Top-10   & 0.693 & 0.746 & 0.957 & 0.610 & 0.670 & 0.856 \\
\hline
\end{tabular}
}
\caption{Results of comparing VSIDS and TEC.}
\label{table:tec}
\vspace{-0.25in}
\end{table}

It is commonly believed that VSIDS focuses on the ``most constrained
part of the formula"~\cite{manysat}, and that this is responsible for
its effectiveness. However, the term ``most constrained part of the
formula'' has previously not been well-defined in a mathematically
precise manner. One intuitive way to define the constrainedness of a
variable is to analyze the Boolean formula, and count how many clauses
a variable occurs in. The variables can then be ranked based on this
measure. In fact, this measure is the basis of the branching heuristic
called DLIS~\cite{dlis}, and was once the dominant branching heuristic
in SAT solvers.  We show that graph centrality measures are a good way
of mathematically defining this intuitive notion of syntactic
``constrainedness of variables'' that has been used by the designers
of branching heuristics. Degree centrality of a vertex in the VIG is
indeed equal to the number of clauses it belongs to, hence it is a
good basis for guessing the constrained variables for the same
reason. Eigenvector centrality extends this intuition by further
increasing the ranks of variables close in proximity to other
constrained variables in the VIG. Additionally, as the dynamic
structure of the VIG evolves due to the addition of learnt clauses by
the solver, the most highly constrained variables in a given instance
also change over time. Hence we incorporated learnt clauses and
temporal information into the TVIG to account for changes in
variables' constrainedness over time.

Besides the success of branching heuristics like VSIDS and DLIS, there
is additional evidence that the syntactic structure is
important for making good branching decisions. For example, Iser et
al.  discovered that initializing the VSIDS activity based on
information computed on the abstract syntax tree of their translator
has a positive impact on solving time~\cite{kodkodconstrainedness}. In
a different paper~\cite{satcommunitypaper}, the authors have shown
that the graph-theoretic community structure strongly influences the
running time of CDCL SAT solvers. This is more evidence of how CDCL
SAT solver performance is influenced by syntactic graph properties of
input formulas. Finally, by combining the results of this section with
Contribution I, we conclude that VSIDS picks
{\it high-centrality bridge variables}.

\section{Contribution IV: Exponential Moving Average and Multiplicative Decay}
\label{sec:ema}

In this section, we argue that the multiplicative decay aspect of the
VSIDS branching heuristic is a form of exponential moving average
(EMA)~\cite{exponentialsmoothing}.  It is the inclusion of multiplicative decay in VSIDS that gives
it its distinctive feature of focusing its search based on recent
conflicts. The original Chaff paper~\cite{chaff} and
patent~\cite{chaffpatent} rather cryptically mentioned that VSIDS acts
like a ``low-pass filter''. They do not specify what signals are being
fed to this filter, and why the high-frequency components are being
filtered out and discarded.

In his paper~\cite{armin2008}, Armin Biere was perhaps the first to
articulate the idea that additive bumping of variable scores can be
viewed as a signal (a square wave, to be more precise) over the run of
the solver. More precisely, at every time step, the signal of a variable is 1
if it is bumped, or 0 otherwise. Armin Biere formalized
\emph{normalized VSIDS}~\cite{armin2008} as $ s_n= (1 - f) \times
\sum_{k=1}^n \delta_k \times f^{n-k} $.  $s_n$ is the normalized VSIDS
activity of a variable $v$ after the $n^{th}$ conflict. $\delta_k = 1$
if variable $v$ was bumped in the $k^{th}$ conflict, otherwise
$\delta_k = 0$. $f$ is the decay factor.

While Huang et al.~\cite{zhang2012} referred to VSIDS as an EMA, we
will show this explicitly. We not only characterize VSIDS as an EMA
explicitly, but also describe why this is crucial to the effectiveness
of VSIDS as a branching heuristic. In the next section we leverage
this connection between EMA and VSIDS to propose an adaptive VSIDS
branching heuristic inspired by an adaptive version of EMA.

EMA is a form of exponential smoothing, used in getting rid of noise
(variables whose VSIDS scores are akin to high-frequency signals) in
time series data (the signals due to VSIDS scores). Exponential
smoothing is a class of techniques to mitigate the effect of random
noise in time series data for the purpose of analysis and
forecasting. Armin Biere's normalized VSIDS equation can be rewritten
to the following recursive formula: $ s_n = (1 - f) \times \delta_n +
f \times s_{n - 1} $.  This formula fits exactly the definition of
Brown's simple exponential smoothing, also known as exponential moving
average.  Therefore normalized VSIDS is exactly an EMA over the
$\delta$ time series.
The EMA causes VSIDS to favor variables that ``persistently''
occur in ``recent'' conflicts. A rationale why this is effective could
be as follows: A conflict essentially points to faulty judgment by
the solver in assigning values to variables. If a set of variables are
at the root of a faulty judgment and thus occurs in a conflict,
then they would repeatedly occur in related faulty judgments and
hence in related conflicts. Variables that occur persistently in
``recent'' conflicts could be a good guess for the root cause of those
conflicts. Hence, perhaps the most effective search strategy is to
focus on determining this root cause. The learnt clauses that result
from such a strategy improve in quality with time, until such time
that the root cause of a set of faulty judgment has been determined
and enshrined as a learnt clause.

\section{Contribution V: A Faster Branching Heuristic Based On Adaptive Moving Average}
\label{sec:dvsids}

In this section, we report on our design of a better VSIDS based on
the knowledge that VSIDS decay is a form of EMA. The EMA is integral
to VSIDS performance as a branching heuristic, and now that the
connection between EMA and VSIDS is established, all the literature on
EMA and other time series data analysis are directly applicable to
VSIDS.

\vs
\noindent{\bf Adaptive Moving Average.} Given that VSIDS decay is a
form of EMA, we studied the literature of EMA from the financial
domain~\cite{kaufmanama}, where it is known that the fixed decay factor can be undesirable. A moving average with a large decay factor would lag behind
fast moving markets whereas a small decay factor would fail to smooth out a
lot of noise.  Kaufman~\cite{kaufmanama} noted that a fixed decay
factor performs poorly when the market volatility changes. He devised
\emph{adaptive moving average} where the decay factor (also known as
smoothing constant) is determined by the market volatility to minimise
lag and noise.  By fluctuating the decay factor when necessary,
adaptive moving average is better than EMA at uncovering trends in the
market.

Just like how markets can go up and down, a CDCL SAT solver can go up
and down in ``productivity'' over time.  For example, Audemard and
Simon~\cite{glucose} discovered that a learnt clause with lower
literals blocks distance (LBD)~\cite{glucose} is of higher quality. LBD of a clause
is defined to be the number of decision levels that its variables span.
If
the solver is in a search space that produces many learnt clauses with low
LBD, then we want to encourage the solver to stay within that search
space.  We do so by adjusting the VSIDS decay factor to be closer to
1, i.e., decay slower. On the other hand, if the solver is in a search
space that produces many learnt clauses with high LBD, it is best to choose a
smaller decay factor, i.e., decay faster.  Based on this insight, we
devised a new VSIDS heuristic called \dvsids by extending \mvsids with
an adaptive moving average.  \dvsids maintains a floating-point number
\texttt{lbdema} equal to the exponential moving average of the learnt
clause LBDs.  \texttt{lbdema} is updated after every learnt clause and
this number will be used to adjust the decay factor of the variables'
activities. In \mvsids, the variables' activities are decayed by
multiplying with a constant decay factor, typically 0.95, after each
conflict. Whereas in \dvsids, the decay factor is adjusted based on
the LBD of the learnt clause.  If the LBD of the learnt clause is
greater than \texttt{lbdema}, then use a decay factor of 0.75,
otherwise use a decay factor of 0.99.  Our website has all the code.

\vs
\noindent{\bf Experimental Setup and Methodology.} The experiments
were performed on the application and combinatorial categories of the SAT
Competition 2013. For each instance with a timeout of 5000 seconds as per
competition rules, we ran an
unmodified MiniSAT 2.2.0 and a modified MiniSAT 2.2.0 with \dvsids
on StarExec~\cite{starexec}.

\vs
\noindent{\bf Results and Interpretations.}  Our
\dvsids solved 351 instances whereas \mvsids solved 343 instances, an
increase of 2.4\% more solved instances.

\section{Interpretation of Results}
\label{sec:interpret}

We began our research by posing a series of questions regarding VSIDS,
and we now interpret the results obtained in light of these questions.

\vs
\noindent{\bf What is special about the class of variables that VSIDS
  chooses to additively bump?} (Answered by Contributions I and III.)
In the bridge variables experiment (Section~\ref{sec:focus}), we
showed that VSIDS disproportionately favored bridge variables.
Even though SAT instances have large number of bridge
variables on average, the frequency with which VSIDS picks, bumps, and
learns bridge variables is much higher. There is no a priori reason to
believe that VSIDS would behave like this. This surprising result,
plus a previous result that good community structure correlates with
faster solving time~\cite{satcommunitypaper}, suggests CDCL solvers
exploit community structure. More precisely, they target variables
linking distinct communities, possibly as a way to solve by
divide-and-conquer approach.

In the VSIDS vs. TGC experiments (Section~\ref{sec:resultstgc}), we
used the Spearman's rank correlation coefficient to show that the
VSIDS and TGC rankings are \emph{strongly correlated}.  From our
experiments, we can say that for all the VSIDS variants considered in
this paper, additive bumping matches with the increase in centrality of
the chosen variables. We also observe from our results that the
variables that solvers pick for branching have very high TGC rank.
The concept of centrality allows us to define in a
mathematically precise the intuition many solver developers have had,
i.e., that branching on ``highly constrained variables'' is an
effective strategy. Our bridge variable experiment combined with the
TGC experiment suggests that VSIDS focuses on {\it high-centrality
  bridge variables}.

\vs
\noindent{\bf What role does multiplicative decay play in making VSIDS
  so effective?}  (Answered by Contribution IV, that in turn led to a
new adaptive VSIDS presented as Contribution V.) We show that
multiplicative decay is essentially a form of exponential smoothing
(Section~\ref{sec:ema}).  We add an explanation as to why this is
important, namely, that exponential smoothing favors variables that
persistently occur in conflicts and this is a better strategy for
root-cause analysis. We designed a new VSIDS technique, we call
\dvsids, based on the above results, wherein we rapidly decay the
VSIDS activity if the learnt clause LBDs are
large~(Section~\ref{sec:dvsids}). We showed that this technique is
better than \mvsids on the SAT Competition 2013 benchmark.

\vs
\noindent{\bf Is VSIDS temporally and spatially focused?}  (Answered
by Contribution II.) We show that VSIDS exhibits \textit{spatial
  focus} and \textit{temporal focus} (Section~\ref{sec:focus}), forms
of locality in search. While there has been speculation among solver
researchers that that CDCL with VSIDS solvers perform local search, we
precisely define spatial and temporal locality in terms of the
community structure.

\section{Related Work}
\label{sec:related}

Marques-Silva and Sakallah are credited with inventing the CDCL
technique~\cite{grasp}.  The original VSIDS heuristic was invented by
the authors of Chaff~\cite{chaff}. Armin Biere~\cite{armin2008} described
the low-pass filter behavior of VSIDS, and Huang et
al.~\cite{zhang2012} stated that VSIDS is essentially an
EMA. Katsirelos and Simon~\cite{eigensat} were the first to publish a
connection between eigenvector centrality and branching heuristics. In
their paper~\cite{eigensat}, the authors computed eigenvector
centrality (via Google PageRank) only once on the original input
clauses and showed that most of the decision variables have higher
than average centrality. Also, it bears stressing that their
definition of centrality is not temporal. By contrast, our results
correlate VSIDS ranking with temporal degree and eigenvector
centrality, and show the correlation holds dynamically throughout the
run of the solver. Also, we noticed that the correlation is also
significantly stronger after extending centrality with temporality.
Simon and Katsirelos do hypothesize that VSIDS may be picking bridge
variables (they call them fringe variables). However, they do not
provide experimental evidence for this. To the best of our knowledge,
we are the first to establish the following results regarding VSIDS:
first, VSIDS picks, bumps, and learns high-centrality bridge
variables; second, VSIDS-influenced search is more spatially and
temporally focused than other branching heuristics we considered;
third, explain the importance of EMA (multiplicative decay) to the
effectiveness of VSIDS; and fourth, invent a new adaptive VSIDS
branching heuristic based on our observations.

\section{Conclusions and Future Work}

In this paper we present various empirically-verified findings on
VSIDS. We show that VSIDS tends to favor the high-centrality bridge
variables in the community structure of the Boolean formula. In
addition, we show that VSIDS focuses on a small subset of communities
in the graph of a SAT instance during search. Lastly, we explain the
multiplicative decay of VSIDS with EMA and use this finding to devise
a new branching heuristic we call \dvsids.  These results put together
show that community structure, graph centrality, and exponential
smoothing are important lenses through which to understand the
behavior of the VSIDS family of branching heuristics and CDCL SAT
solving. In the future, we plan to strengthen our results by
considering a larger number of benchmarks, solvers, branching
heuristics, and graph representations.

\section{Acknowledgement}
We thank Kaveh Ghasemloo for his help in refining our TGC model and for his insight on the connection between VSIDS decay and exponential moving average.


\bibliographystyle{splncs03}
\bibliography{bibliography}

\begin{thebibliography}{10}
\providecommand{\url}[1]{\texttt{#1}}
\providecommand{\urlprefix}{URL }

\bibitem{starexec}
Starexec, \url{http://www.starexec.org/}

\bibitem{satconf}
Proceedings of {P}ast {SAT} {C}onferences (2013),
  \url{http://www.satisfiability.org}

\bibitem{satcomp}
{SAT} {C}ompetition {W}ebsite (2013), \url{http://www.satcompetition.org}

\bibitem{sharcnet}
{SHARCNET} {W}ebsite (2013), \url{https://www.sharcnet.ca}

\bibitem{atserias}
Atserias, A., Fichte, J.K., Thurley, M.: Clause-learning algorithms with many
  restarts and bounded-width resolution. In: Theory and Applications of
  Satisfiability Testing-SAT 2009, pp. 114--127. Springer (2009)

\bibitem{glucose}
Audemard, G., Simon, L.: Glucose: a solver that predicts learnt clauses
  quality. IJCAI  9,  399--404 (2009)

\bibitem{beame}
Beame, P., Kautz, H.A., Sabharwal, A.: Towards understanding and harnessing the
  potential of clause learning. Journal of Artificial Intelligence Research
  (JAIR)  22,  319--351 (2004)

\bibitem{armin2008}
Biere, A.: Adaptive restart strategies for conflict driven {SAT} solvers. In:
  Proceedings of the 11th International Conference on Theory and Applications
  of Satisfiability Testing. pp. 28--33. SAT'08, Springer-Verlag, Berlin,
  Heidelberg (2008)

\bibitem{lingeling}
Biere, A.: Lingeling (2010)

\bibitem{louvain}
Blondel, V.D., Guillaume, J.L., Lambiotte, R., Lefebvre, E.: Fast unfolding of
  communities in large networks. Journal of Statistical Mechanics: Theory and
  Experiment  2008(10),  P10008 (2008)

\bibitem{exponentialsmoothing}
Brown, R.G.: Exponential Smoothing for predicting demand. Little (1956)

\bibitem{bohm}
Buro, M., B{\"u}ning, H.K.: Report on a {SAT} competition. Fachbereich
  Math.-Informatik, Univ. Gesamthochschule (1992)

\bibitem{clauset2004finding}
Clauset, A., Newman, M.E., Moore, C.: Finding community structure in very large
  networks. Physical review E  70(6),  066111 (2004)

\bibitem{npcook}
Cook, S.A.: The complexity of theorem-proving procedures. In: Proceedings of
  the Third Annual ACM Symposium on Theory of Computing. pp. 151--158. STOC
  '71, ACM, New York, NY, USA (1971)

\bibitem{minisat}
Een, N., S{\"o}rensson, N.: {M}ini{S}at: {A} {SAT} solver with conflict-clause
  minimization. Sat  5 (2005)

\bibitem{centralitysocial}
Faust, K.: Centrality in affiliation networks. Social networks  19(2),
  157--191 (1997)

\bibitem{fishertransformation}
Fisher, R.A.: Frequency distribution of the values of the correlation
  coefficient in samples from an indefinitely large population. Biometrika pp.
  507--521 (1915)

\bibitem{mom}
Freeman, J.W.: Improvements to Propositional Satisfiability Search Algorithms.
  Ph.D. thesis, Philadelphia, PA, USA (1995), uMI Order No. GAX95-32175

\bibitem{centralityfreeman1979}
Freeman, L.: Centrality in social networks conceptual clarification. Social
  Networks  1(3),  215--239 (1979)

\bibitem{gini}
Gini, C.: Measurement of inequality of incomes. The Economic Journal pp.
  124--126 (1921)

\bibitem{community}
Girvan, M., Newman, M.E.: Community structure in social and biological
  networks. Proceedings of the National Academy of Sciences  99(12),
  7821--7826 (2002)

\bibitem{temporalsocialnetwork1}
Gloor, P., Krauss, J., Nann, S., Fischbach, K., Schoder, D.: Web science 2.0:
  Identifying trends through semantic social network analysis. In:
  Computational Science and Engineering, 2009. CSE '09. International
  Conference on. vol.~4, pp. 215--222 (Aug 2009)

\bibitem{poweriteration}
Golub, G.H., Van~Loan, C.F.: Matrix computations, vol.~3. JHU Press (2012)

\bibitem{manysat}
Hamadi, Y., Jabbour, S., Sais, L.: Many{SAT}: a parallel {SAT} solver. JSAT
  6(4),  245--262 (2009)

\bibitem{sls}
Hoos, H.H., St{\"u}tzle, T.: Stochastic Local Search: Foundations \&
  Applications. Morgan Kaufmann Publishers Inc., San Francisco, CA, USA (2004)

\bibitem{zhang2012}
Huang, R., Chen, Y., Zhang, W.: {SAS}+ planning as satisfiability. J. Artif.
  Int. Res.  43(1),  293--328 (Jan 2012)

\bibitem{kodkodconstrainedness}
Iser, M., Taghdiri, M., Sinz, C.: Optimizing {M}ini{SAT} variable orderings for
  the relational model finder {K}odkod. In: Proceedings of the 15th
  International Conference on Theory and Applications of Satisfiability
  Testing. pp. 483--484. SAT'12, Springer-Verlag, Berlin, Heidelberg (2012)

\bibitem{jeroslowwang}
Jeroslow, R.G., Wang, J.: Solving propositional satisfiability problems. Annals
  of mathematics and Artificial Intelligence  1(1-4),  167--187 (1990)

\bibitem{empircalsat}
Katebi, H., Sakallah, K.A., Marques-Silva, J.P.: Empirical study of the anatomy
  of modern {SAT} solvers. In: Proceedings of the 14th International Conference
  on Theory and Application of Satisfiability Testing. pp. 343--356. SAT'11,
  Springer-Verlag, Berlin, Heidelberg (2011)

\bibitem{eigensat}
Katsirelos, G., Simon, L.: Eigenvector centrality in industrial {SAT}
  instances. In: Milano, M. (ed.) Principles and Practice of Constraint
  Programming, pp. 348--356. Lecture Notes in Computer Science, Springer Berlin
  Heidelberg (2012)

\bibitem{kaufmanama}
Kaufman, P.J.: Trading systems and methods. John Wiley \& Sons (2013)

\bibitem{zchaff}
Mahajan, Y.S., Fu, Z., Malik, S.: Zchaff2004: An efficient {SAT} solver. In:
  Proceedings of the 7th International Conference on Theory and Applications of
  Satisfiability Testing. pp. 360--375. SAT'04, Springer-Verlag, Berlin,
  Heidelberg (2005)

\bibitem{dlis}
Marques-Silva, J.P.: The impact of branching heuristics in propositional
  satisfiability algorithms. In: Progress in Artificial Intelligence, pp.
  62--74. Springer (1999)

\bibitem{grasp}
Marques-Silva, J.P., Sakallah, K.A.: Grasp: A search algorithm for
  propositional satisfiability. Computers, IEEE Transactions on  48(5),
  506--521 (1999)

\bibitem{chaffpatent}
Moskewicz, M.W., Madigan, C.F., Malik, S.: Method and system for efficient
  implementation of boolean satisfiability (Aug~26 2008), uS Patent 7,418,369

\bibitem{chaff}
Moskewicz, M.W., Madigan, C.F., Zhao, Y., Zhang, L., Malik, S.: Chaff:
  Engineering an efficient {SAT} solver. In: Proceedings of the 38th Annual
  Design Automation Conference. pp. 530--535. DAC '01, ACM, New York, NY, USA
  (2001)

\bibitem{nadel2010assignment}
Nadel, A., Ryvchin, V.: Assignment stack shrinking. In: Theory and Applications
  of Satisfiability Testing--SAT 2010, pp. 375--381. Springer (2010)

\bibitem{satcommunitypaper}
Newsham, Z., Ganesh, V., Fischmeister, S., Audemard, G., Simon, L.: Impact of
  community structure on {SAT} solver performance. In: Theory and Applications
  of Satisfiability Testing--SAT 2014, pp. 252--268. Springer (2014)

\bibitem{knot}
Pipatsrisawat, K., Darwiche, A.: On the power of clause-learning {SAT} solvers
  with restarts. In: Principles and Practice of Constraint Programming-CP 2009,
  pp. 654--668. Springer (2009)

\bibitem{spearman}
Spearman, C.: The proof and measurement of association between two things. The
  American journal of psychology  15(1),  72--101 (1904)

\bibitem{gouldindexjustification}
Straffin, P.D.: Linear algebra in geography: Eigenvectors of networks.
  Mathematics Magazine  53(5),  269--276 (1980)

\bibitem{timedpagerank}
Yu, P.S., Li, X., Liu, B.: Adding the temporal dimension to search - a case
  study in publication search. In: Skowron, A., Agrawal, R., Luck, M.,
  Yamaguchi, T., Morizet-Mahoudeaux, P., Liu, J., Zhong, N. (eds.) Web
  Intelligence. pp. 543--549. IEEE Computer Society (2005)

\bibitem{zhang2013online}
Zhang, W., Pan, G., Wu, Z., Li, S.: Online community detection for large
  complex networks. In: Proceedings of the Twenty-Third international joint
  conference on Artificial Intelligence. pp. 1903--1909. AAAI Press (2013)

\end{thebibliography}

\newpage
\appendix
\section{Appendix}

\begin{figure}[h]
	\setlength{\abovecaptionskip}{1pt}
	
	\setlength{\belowcaptionskip}{1pt}
        \minipage{\textwidth}
        \includegraphics[width=\linewidth]{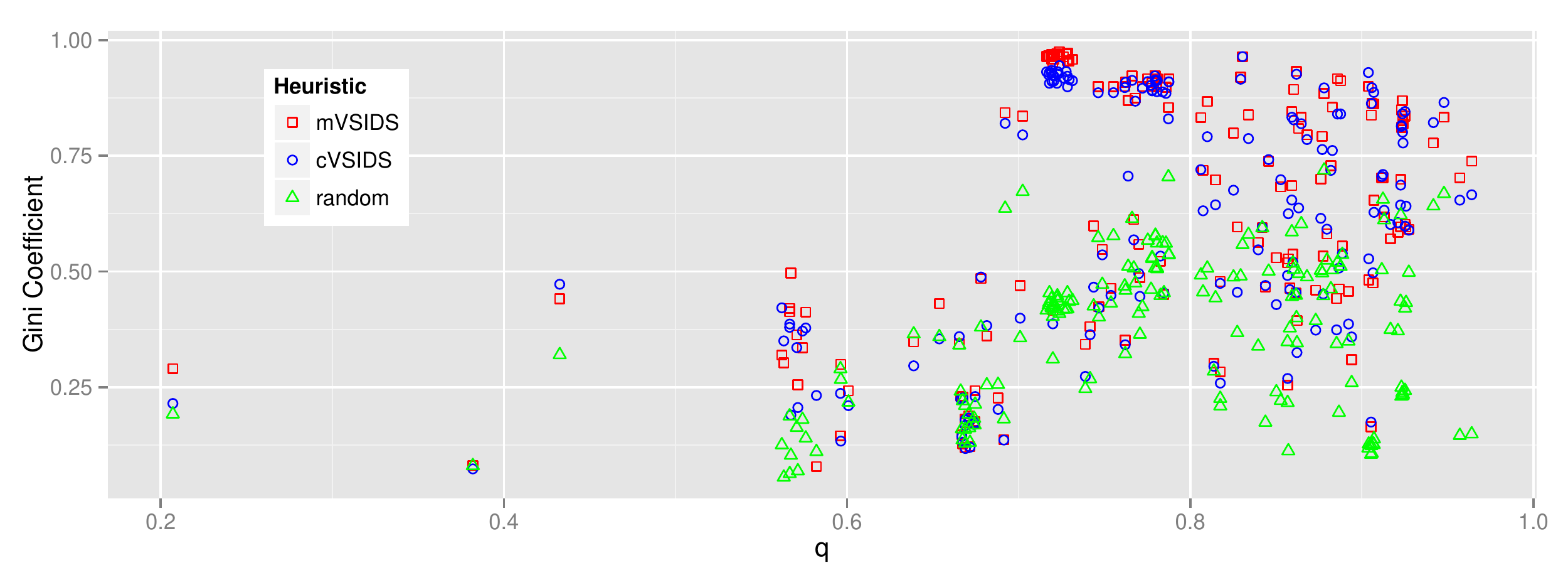}
        \caption*{(a) Application}
        \endminipage\hfill
        
        \minipage{\textwidth}
        \includegraphics[width=\linewidth]{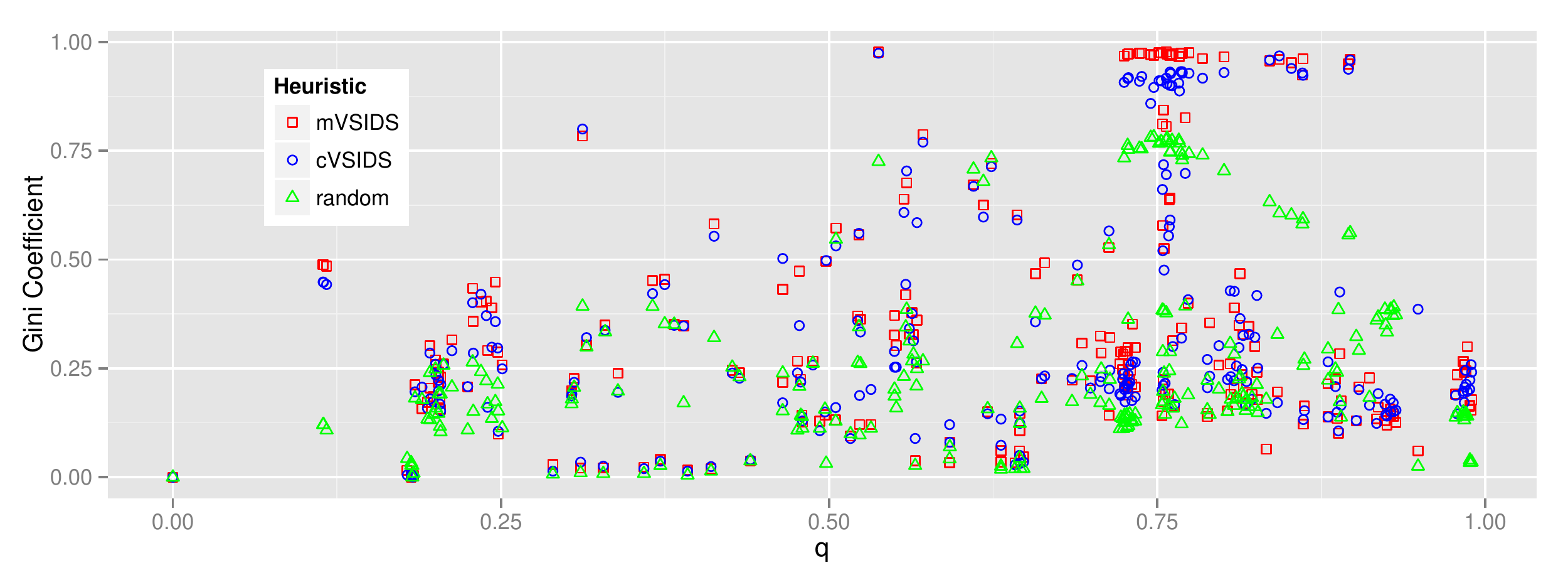}
        \caption*{(b) Combinatorial}
        \endminipage\hfill
        
        \minipage{\textwidth}%
        \includegraphics[width=\linewidth]{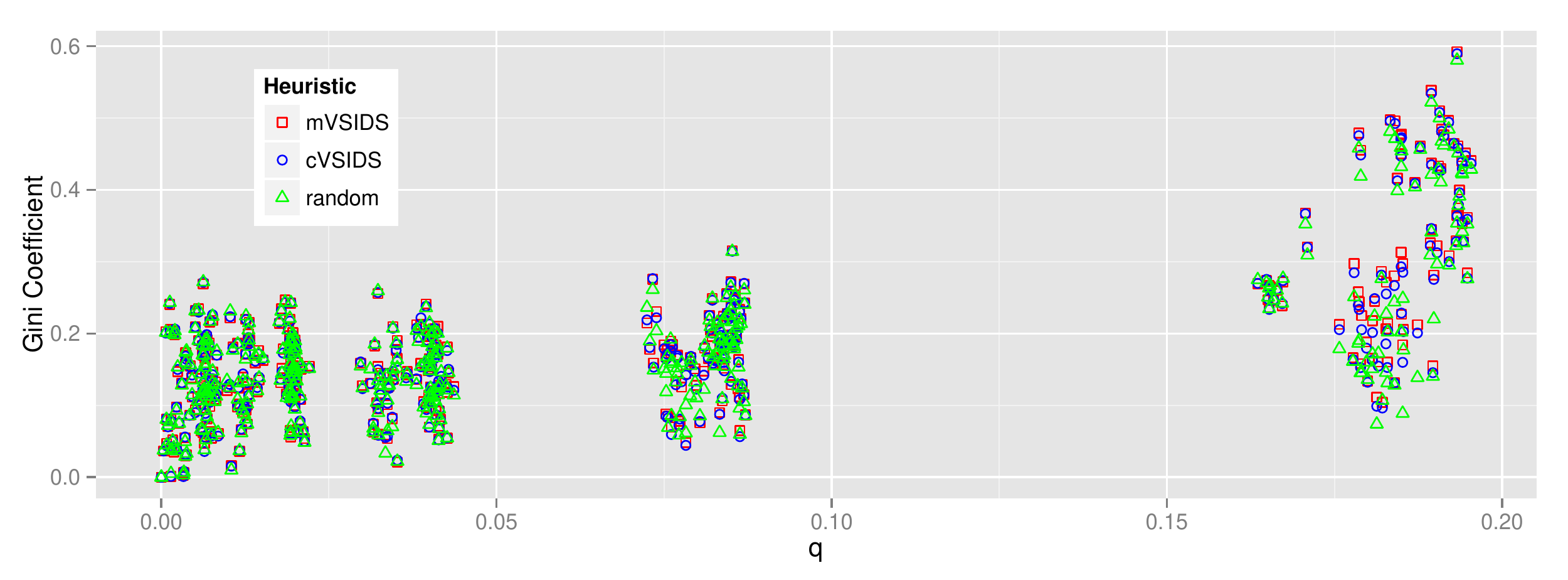}
        \caption*{(c) Random}
        
        \endminipage
        \caption{ VSIDS heuristics are much more spatially focused than random branching. For each instance and each branching heuristic, we plot the Gini coefficient of the normalized community hits (as in Section \ref{sec:focus}). Higher points indicate more spatial focus. X-axes denote the modularity of the instance's community structure -- a standard metric for the quality of a community structure (higher is better).}
        \label{spatial-graphs}
        \vspace{-2cm}
\end{figure}

\begin{figure}[h!]
	\setlength{\abovecaptionskip}{1pt}
	
	\setlength{\belowcaptionskip}{1pt}
        \minipage{\textwidth}
        \includegraphics[width=\linewidth]{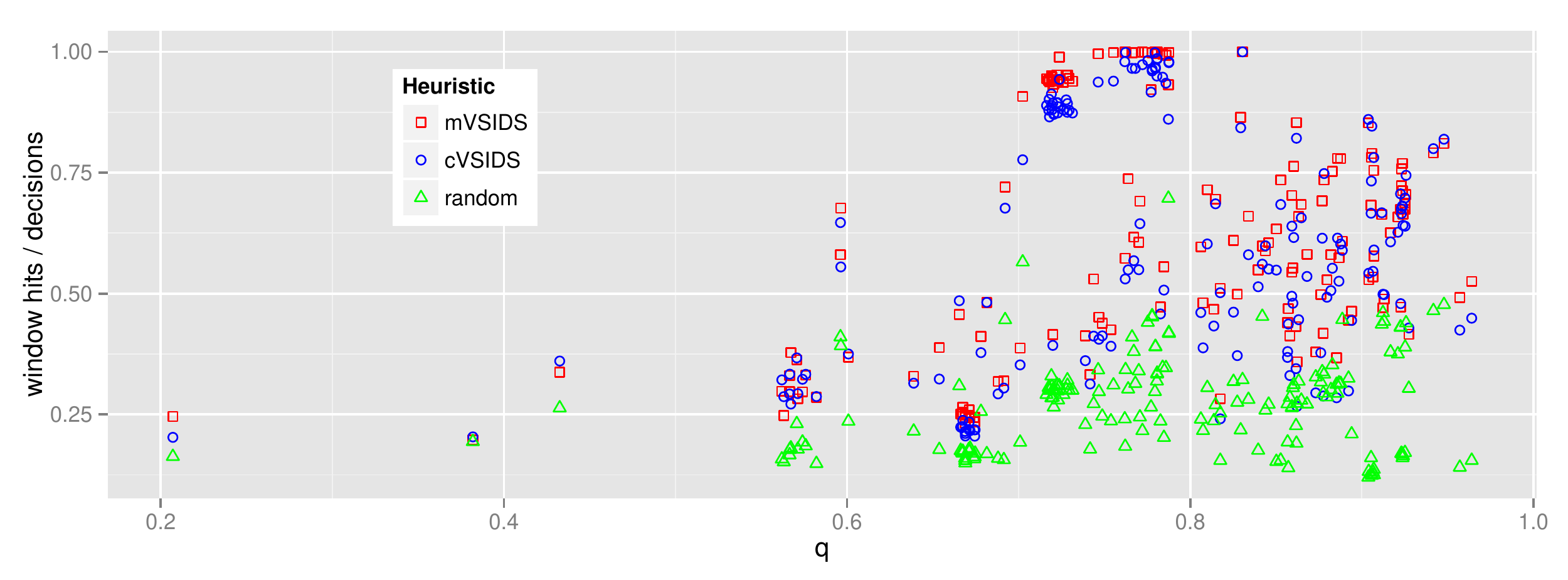}
        \caption*{(a) Application}
        \endminipage\hfill
        
        \minipage{\textwidth}
        \includegraphics[width=\linewidth]{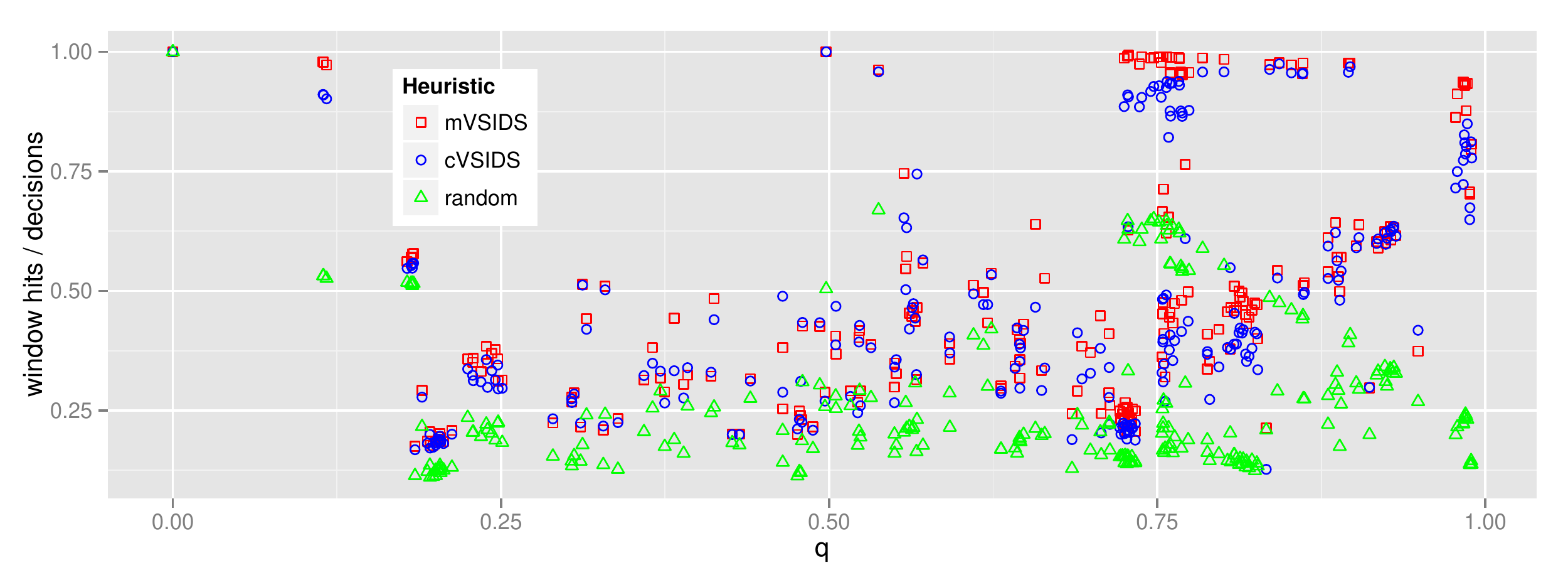}
        \caption*{(b) Combinatorial}
        \endminipage\hfill
        
        \minipage{\textwidth}%
        \includegraphics[width=\linewidth]{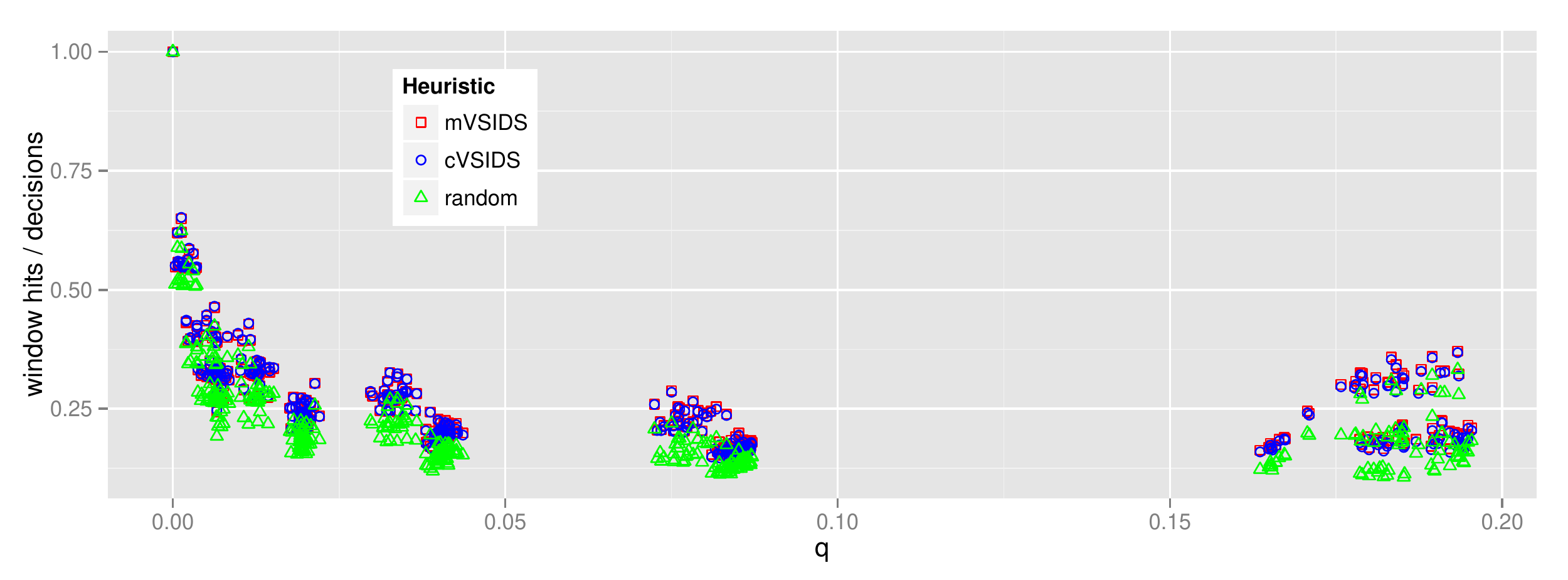}
        \caption*{(c) Random}
        
        \endminipage
        \caption{\tempexp comparing \mvsids, \cvsids, and random, with a window size equal to 10\% of the total number of communities (as in Section \ref{sec:focus}). Higher points indicate better temporal focus. Both VSIDS heuristics significantly dominate random branching, and \mvsids is slightly more focused than \cvsids on average.} 
        \label{temporal-graphs}
\end{figure}

\begin{table}[h]
\begin{tabular}{cc}
\includegraphics[width=0.45\linewidth]{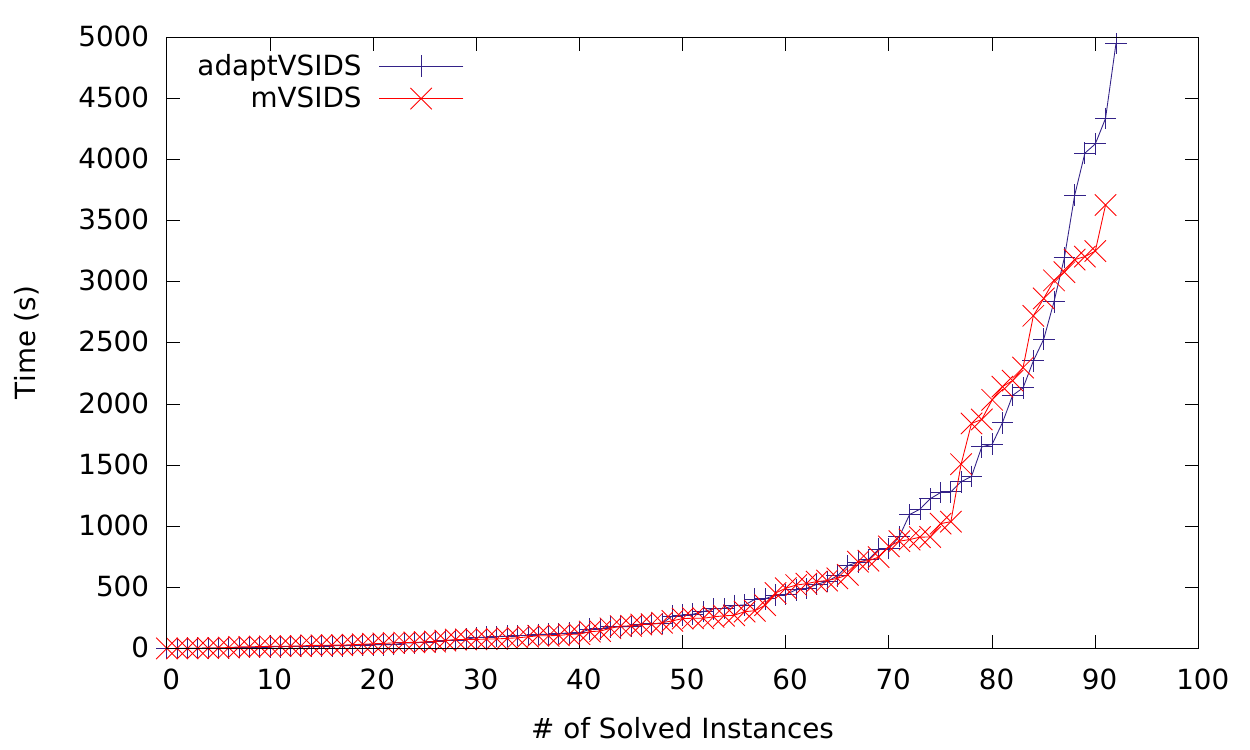} & \includegraphics[width=0.45\linewidth]{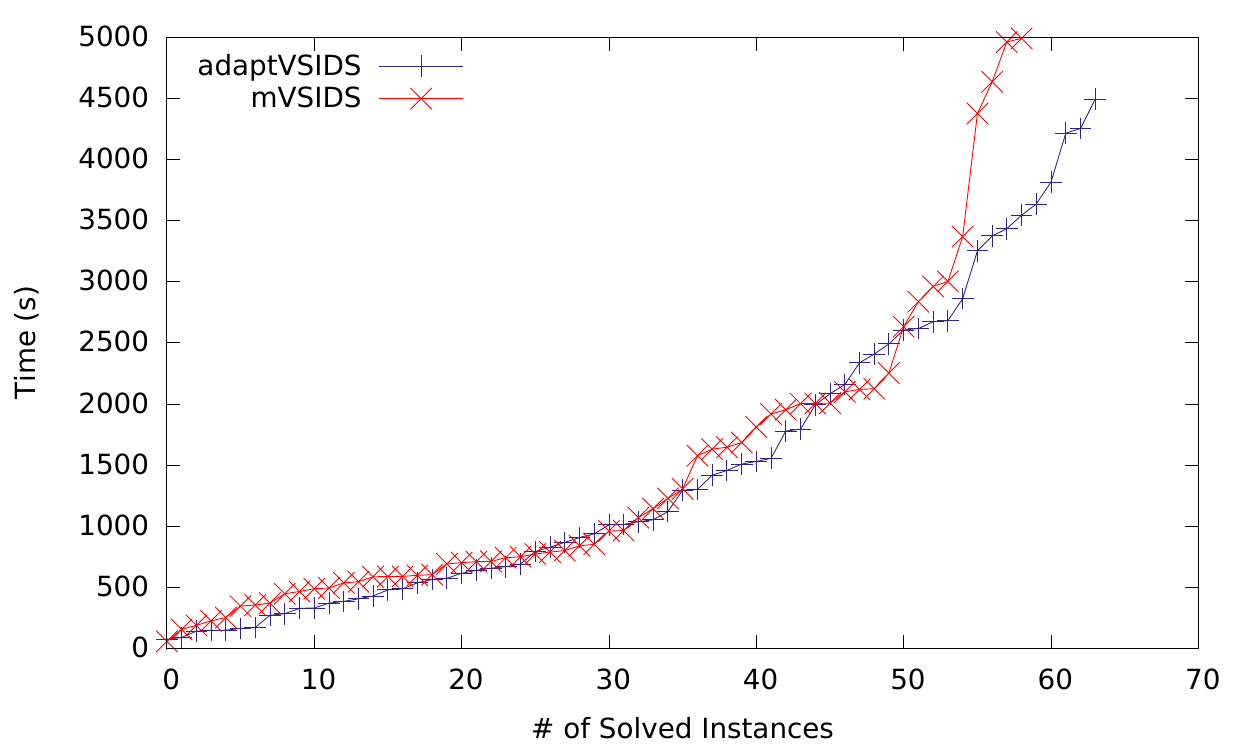} \\
(a) satisfiable application instances & (b) unsatisfiable application instances\\
\\
\includegraphics[width=0.45\linewidth]{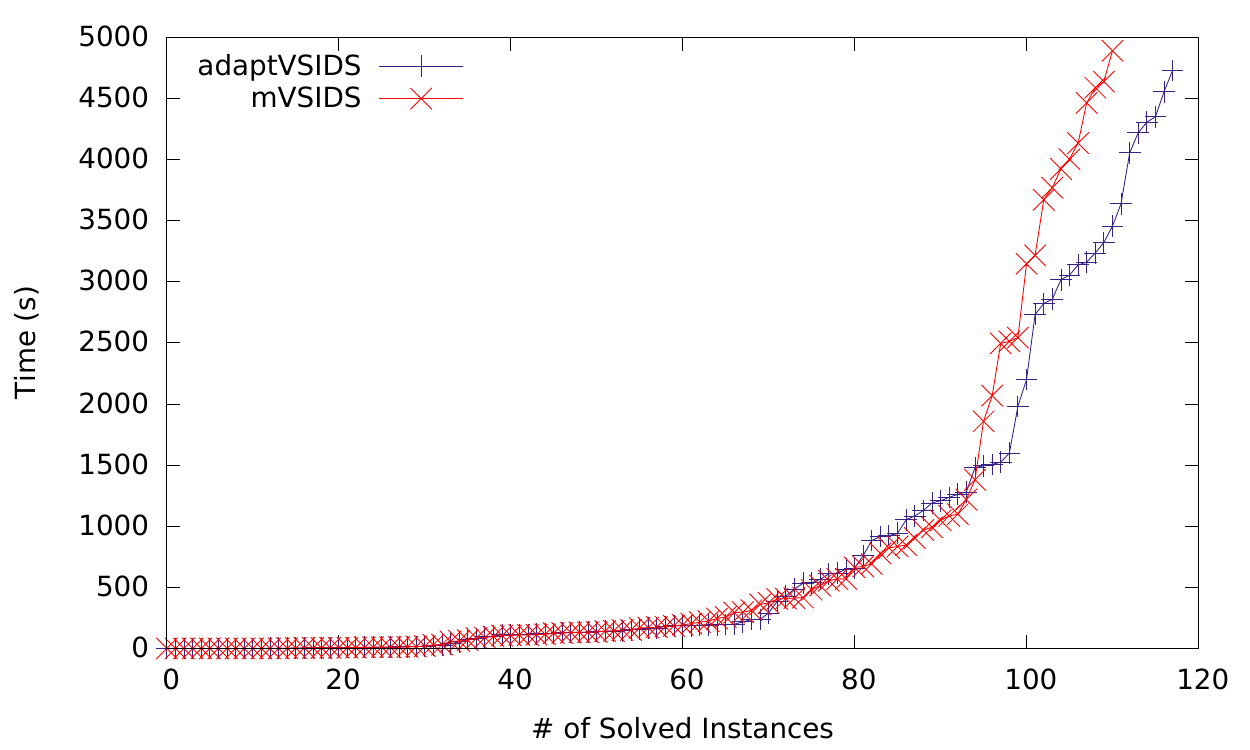} & \includegraphics[width=0.45\linewidth]{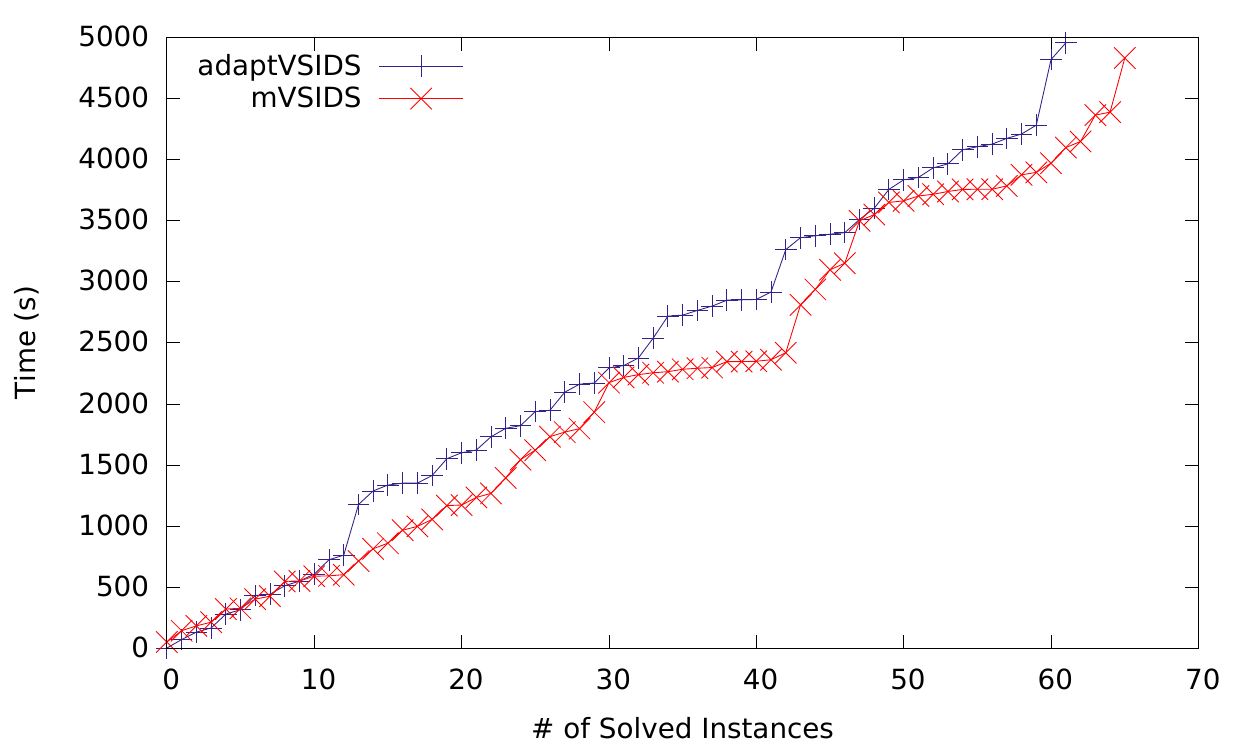} \\
(c) satisfiable combinatorial instances & (d) unsatisfiable combinatorial instances\\
\end{tabular}
\caption{Cactus plots for the \dvsids experiment. The results are split into two categories (application or combinatorial) and two statuses (satisfiable or unsatisfiable). A point (80, 1500) can be interpreted as follows: there are 80 instances that take less than 1500 seconds to solve with the respective branching heuristic.}
\end{table}

\end{document}